\documentclass[12pt,a4paper]{article}
\usepackage[truedimen,margin=30mm]{geometry} 

\usepackage{mathrsfs}
\usepackage{amssymb}
\usepackage{amsmath}
\usepackage{ascmac}
\usepackage{amsthm}
\usepackage[dvips]{graphicx}
\usepackage{natbib}
\usepackage{setspace}
\usepackage{times}
\usepackage{comment}

\usepackage{color}

\usepackage{titlesec}
\titleformat*{\section}{\large\bfseries}
\titleformat*{\subsection}{\it}

\def\ep{{\varepsilon}}

\title{{\bf Noise-Robust Phase Connectivity Estimation via Bayesian Circular Functional Models}\footnote{\today}}

\date{}

\begin{document}

\maketitle
\doublespacing

\vspace{-1.5cm}
\begin{center}
{\large Shonosuke Sugasawa$^{1,2}$, Takeru Matsuda$^{3,2}$ and Tomoyuki Nagakawa$^{4,2}$}

\medskip

\medskip
\noindent
$^1$Faculty of Economics, Keio University\\
$^2$RIKEN Center for Brain Science\\
$^3$Graduate School of Information Science and Technology,
The University of Tokyo\\
$^4$School of Data Science, Meisei University
\end{center}

\vspace{0.5cm}
\begin{center}
{\bf \large Abstract}
\end{center}

The phase locking value (PLV) is a widely used measure to detect phase connectivity. Main drawbacks of the standard PLV are it can be sensitive to noisy observations and does not provide uncertainty measures under finite samples. To overcome the difficulty, we propose a model-based PLV through nonparametric statistical modeling. Specifically, since the discrete time series of phase can be regarded as a functional observation taking values on circle, we employ a Bayesian model for circular-variate functional data, which gives denoising and inference on the resulting PLV values. The proposed model is defined through ``wrapping" functional Gaussian models on real line, for which we develop an efficient posterior computation algorithm using Gibbs sampler. The usefulness of the proposed method is demonstrated through simulation experiments based on real EEG data.


\newpage
\section{Introduction}

Electroencephalography (EEG) is a widely used non-invasive tool for measuring brain activity with high temporal resolution. 
A key aspect of EEG analysis is the investigation of phase synchrony between oscillatory signals from different brain regions, which provides important insights into functional connectivity \citep{stam2007phase,cao2022brain}. 
Among various measures of phase synchrony, the phase locking value (PLV) is one of the most popular due to its simplicity, interpretability, and broad applicability in neuroscience and clinical research \citep{lachaux1999measuring}.
Despite its popularity, the conventional PLV has notable limitations in practice.
First, it is typically computed in a purely descriptive manner from observed phase data, making it highly sensitive to measurement noise \citep{celka2007statistical}, under which the true PLV could be substantially underestimated.  
Second, the standard PLV does not provide any quantification of statistical uncertainty and the assessment of the statistical significance rely on bootstrap methods \citep[e.g.][]{arnulfo2020long} or surrogate data methods \citep[e.g.][]{lachaux1999measuring}.
However, it may fail to capture the statistical significance under noisy EEG observations.

To solve the aforementioned issue, we propose a new statistical framework for estimating PLV from noisy EEG observations. 
Since the instantaneous phase in EEG can be regarded as a functional observation on the unit circle, namely, a circular function defined over time, we employ a statistical models for for multiple circular functions observed over multiple locations and individuals. 
Specifically, our approach models the underlying denoised phase functions via basis expansion and hierarchical priors, while accounting for periodicity through the wrapped Gaussian distribution \citep{mardia1999directional}.  
This model removes idiosyncratic noise components from the observed data to get accurate estimates of PLV, which we call ``model-based PLV". 
To conduct natural quantification of uncertainty, we consider a Bayesian approach to fit the model and the statistical inference can be performed via Markov Chain Monte Carlo (MCMC) algorithm.
The posterior samples of PLV generated via MCMC allows us to compute interval estimation as well as probabilities for PLV exceeding a given threshold. 
This enables uncertainty-aware connectivity analysis and reduces the risk of overconfident conclusions based on noisy PLV estimates.

Regarding functional modeling, multivariate (real-valued) functional models, particular for graphical modeling, are well established \citep[e.g.][]{zhu2016bayesian,liu2025bayesian}, but it cannot be directly applied to the circular functions requires due to periodicity. 
Ignoring periodicity, as in conventional real-valued models, can result in artificial discontinuities and biased estimates of synchrony \citep[e.g.][]{mardia1999directional,jammalamadaka2001topics}.
In multivariate circular data analysis, there are several approaches to construct circular distributions by mapping multivariate Gaussian variables to the unit circle \citep{coles1998inference, jona2012spatial,mastrantonio2016spatio,wang2015joint} and by using multivariate von Mises distributions \citep{mardia2007protein, Mardia2008multivariate, mardia2014some}.
However, such methods do not assume latent functional structures, and to our knowledge, there has been no attempt to develop hierarchical random effects models for multivariate functional observations. 
Moreover, while the wrapping approach used in the proposed is widely used in modeling circular observations \citep[e.g.][]{gottard2024gaussian, ravindran2011bayesian, jona2012spatial}, it has not been adopted for functional modeling of circular data.

This paper is organized as follows: 
In Section~\ref{sec:method}, we introduce the proposed circular functional models and provide a computational algorithm to obtain the posterior distribution of PLV.
Section~\ref{sec:num} illustrates the effectiveness of the proposed method through simulations and real EEG data analysis.
We give concluding remarks in Section~\ref{sec:conc}.

\section{Wrapped Functional Models for EEG Phase}\label{sec:method}

\subsection{Phase synchrony and phase locking value}

Let $\theta_k(t) \in [0, 2\pi)$ denote the instantaneous phase of the $k$th signal at time $t$, for $k = 1, \dots, p$ and $t = t_1, \dots, t_T$, where each $\theta_k(t)$ is defined on the unit circle. 
These phase values may be extracted from observed oscillatory signals via analytic signal representations such as Hilbert or wavelet transformations \citep{cohen2014analyzing}. 
To quantify the synchrony between any pair of signals $(k,k')$, the phase locking value (PLV) between $\theta_k(t)$ and $\theta_{k'}(t)$ is defined as
\begin{equation}\label{eq:PLV}
\mathrm{PLV}_{k,k'} = \left| \frac{1}{T} \sum_{j=1}^{T} \exp\left\{ i \left( \theta_k(t_j) - \theta_{k'}(t_j) \right) \right\} \right|,
\end{equation}
where $i$ is the imaginary unit \citep{lachaux1999measuring}. The PLV takes values in the interval $[0, 1]$, with values close to $1$ indicating strong phase locking (i.e., consistent phase difference) and values near $0$ indicating weak or no phase synchrony between the two signals.
It is common to define phase connectivity among $p$ locations by treating $\mathrm{PLV}_{k,k'}$ values that exceed a certain threshold as indicative of phase locking between locations $k$ and $k'$ \citep[e.g.][]{myers2016seizure}.
Given replicated observations $\theta_k^{(s)}(t)$ for each location $k = 1, \dots, p$ and subject or trial $s = 1, \dots, n$, it is possible to define PLV for each subject $s$ and pair $(k,k')$, defined as ${\rm PLV}_{k,k'}^{(s)}$. 
Then, the overall PLV can be defined as $\overline{{\rm PLV}}_{k,k'}=n^{-1}\sum_{s=1}^n {\rm PLV}_{k,k'}^{(s)}$.

Although PLV (\ref{eq:PLV}) is widely used in neuroscience and related fields, it is typically computed directly from data as a descriptive statistic. 
This approach does not account for noise included in $\theta_k(t)$ and uncertainty of the calculated PLV under finite number of samples. 
As discussed in \cite{celka2007statistical}, PLV is sensitive to noise and estimated PLV values can be significantly smaller than the true value.
Also, the lack of uncertainty quantification prevents rigorous statistical inference based on PLV. For instance, without credible intervals or hypothesis testing procedures, it is difficult to assess whether observed differences in PLV across experimental conditions are statistically significant or merely due to sampling variability. 
These issues motivate using a model-based version of PLV to conduct denoising and uncertainty quantification.

\subsection{Functional models and wrapping}
In practice, we typically observed a noisy version of $\theta_k^{(s)}(t)$. 
Let $\{Y_k^{(s)}(t)\}_{k=1,\ldots,p}$ denotes a set of $p$ random functions on $t\in \mathcal{T}$ for the $s$th subject, whose value takes on the circle, namely $Y_k^{(s)}(t)\in [0,2\pi)$. 
To define the model for $Y_k^{(s)}(t)$, we first introduce functional random variable, $W_k^{(s)}(t)$, on the real line, namely, $W_k^{(s)}(t)\in \mathcal{R}$.
We then consider the following model for $W_k^{(s)}(t)$:
\begin{equation}\label{eq:model1}
W_k^{(s)}(t)=\sum_{l=1}^L a_{kl}^{(s)}B_l(t) + \ep_k^{(s)}(t), \ \ \ \ k=1,\ldots,p, 
\end{equation}
where $\{B_1(t),\ldots,B_L(t)\}$ is a set of basis functions, $a_{kl}^{(s)}$ is an unknown coefficient, and $\ep_k^{(s)}(t)$ is an measurement error.
We assume that $\ep_k^{(s)}(t)$ and $\ep_{k'}^{(s')}(t')$ are independent when either $k\neq k'$, $t\neq t'$ or $s\neq s'$ holds, and the marginal distribution is $\ep_k^{(s)}(t)\sim N(0, \sigma^2)$.
For sufficiently large $L$, the model (\ref{eq:model1}) provides flexible representation of the underlying (denoised) function.
In our numerical experiments, we employ the P-spline basis function, $B_l(t)=t^{l-1}$ for $l=1,\ldots,q+1$ and $B_{q+k}(t)=(t-\kappa_k)_+^q$ for $k=1,\ldots,K$, where $q$ is the degree of polynomial functions and $\kappa_1,\ldots,\kappa_L$ are a set of fixed knots. 
Note that $L=q+K+1$ in this case. 
The knots can be specified to sufficiently spread the domain $\mathcal{T}$ to achieve flexible representation of the underlying function.

Using $W_k^{(s)}(t)$, we then define the model for the circular-variate functional object, $Y_k^{(s)}(t)$, as 
$$
Y_k(t) = W_k(t) \mod 2\pi.  
$$
for each $t\in \mathcal{T}$. 
Suppose the function value is observed at the discrete points, $t_1,\ldots,t_T$ on $\mathcal{T}$, and define $Y_{kj}^{(s)}\equiv Y_k^{(s)}(t_j)$ and $B_{lj}\equiv B_l(t_j)$.
Then, given the individual-specific coefficient $A_k^{(s)}=(a_{k1}^{(s)},\ldots,a_{kL}^{(s)})$, the density of  $Y_{kj}^{(s)}$ is expressed as 
\begin{equation}\label{eq:wrapped-normal}
f(Y_{kj}^{(s)}) = \frac{1}{\sqrt{2\pi\sigma^2}} \sum_{m=-\infty}^{\infty} \exp\left\{-\frac{(Y_{kj}^{(s)} - \sum_{l=1}^L a_{kl}^{(s)}B_{lj}+ 2\pi m)^2}{2\sigma^2}\right\}.
\end{equation}
The above density is known as the wrapped normal distribution \citep{mardia1999directional}.
For the individual-specific coefficient, $a_{kl}^{(s)}$, we assume a hierarchical random effect structure
$$
a_{kl}^{(s)}|\mu_{kl}\sim N(\mu_{kl}, \tau_l^2), \ \ \ \ 
\mu_{kl}\sim N(\beta_l, \gamma_l^2),
$$
where $\tau_l$, $\beta_l$ and $\gamma$ are unknown parameters. 
Note that we assume conditional independence among the individual-specific coefficients, $a_{kl}^{(1)},\ldots, a_{kl}^{(n)}$ given $\mu_{kl}$, but they are marginally correlated through the common $\mu_{kl}$, which can represent possible heterogeneity of phase curves among individuals. 
Similarly, while the location-specific coefficients, $\mu_{1l},\ldots, \mu_{pl}$ are conditionally independent given $\beta_l$ and $\gamma_l^2$, the mean and variance are different over $l$. 
Under the above random effect structure, the overall prior expectation (grand mean) of the underlying function on real line is $\sum_{l=1}^L \beta_l B_l(t)$ for all $k=1,\ldots,p$ and $s=1,\ldots,n$, which indicates that ${\rm PLV}_{k,k'}$ is exactly equal to $1$. 
Hence, the parameters $\tau_l$ and $\gamma_l$ captures the variability of the underlying functions across different dimensions and subjects, allowing the model to represent situations where PLV takes values less than 1. 
In this way, deviations from the mean function are interpreted as subject- or dimension-specific differences in phase. 
Moreover, the random effect structure imposes a shrinkage toward a common functional form, which acts in the direction of increasing phase synchrony (i.e., toward ${\rm PLV}_{k,k'} = 1$), while mitigating the risk of overfitting that may arise due to the high-dimensional parameter space.

\subsection{Bayesian inference}\label{sec:pos}

To fit the model, we employ a Bayesian approach by assigning conditionally conjugate prior distributions, $\beta_l\sim N(A_0, B_0)$, $\tau_l^2\sim {\rm IG}(\nu_\tau, \eta_\tau)$, $\gamma_l^2\sim {\rm IG}(\nu_\gamma, \eta_\gamma)$ and $\sigma^2\sim {\rm IG}(\nu_\sigma, \nu_\sigma)$.
An advantage of the Bayesian approach is that it provides a coherent framework for quantifying uncertainty of model parameters and hence PLV through posterior distributions. 
This allows us to obtain not only point estimates but also credible intervals.
Since the proposed model includes a large number of latent variables and parameters, the general software for Bayesian computation such as \texttt{Stan} \citep{carpenter2017stan} is not directly applicable due to inefficiency. 
Instead, we develop a tailored Gibbs sampling algorithm exploiting the conditional conjugacy of the priors, which enables efficient posterior computation and scalable inference.

To give an efficient posterior computation algorithm, we employ an alternative expression for the density (\ref{eq:wrapped-normal}).
By introducing a latent variable $Z_{kj}^{(s)}$ taking values on $\{0, \pm 1, \pm 2,\ldots\}$.
Then, the density (\ref{eq:wrapped-normal}) can be augmented as 
\begin{equation}\label{eq:joint}
f(Y_{kj}^{(s)},Z_{kj}^{(s)})=\frac{1}{\sqrt{2\pi\sigma^2}}
\exp\left\{-\frac{(Y_{kj}^{(s)} - \sum_{l=1}^L a_{kl}^{(s)}B_{lj}+ 2\pi Z_{kj}^{(s)})^2}{2\sigma^2}\right\}.
\end{equation}
By marginalizing the density (\ref{eq:joint}) with respect to $Z_{kj}^{(s)}$ gives the marginal density (\ref{eq:wrapped-normal}).
Therefore, given the latent variable $Z_{ik}^{(s)}$, the likelihood for the random coefficient $a_{kl}^{(s)}$ is the same as the normal likelihood. 
Based on the representation (\ref{eq:joint}) of the wrapped distribution, we can develop an efficient Markov Chain Monte Carlo algorithm using a Gibbs sampler, where the detailed derivation and sampling steps are deferred to the Appendix.

Given the posterior samples of $a_{kl}^{(s)}$ and $Z_{kl}^{(s)}$, the posterior distribution of the denoised function
$$
\theta_{kj}^{(s)}(t) \equiv 
\sum_{l=1}^L \alpha_{kl}^{(s)}B_{l}(t)-2\pi Z_{kj}^{(s)}
$$
can be generated at arbitrary $t\in \mathcal{T}$.
Then, we can simulate posterior distribution of PLV for each subject $s$ through (\ref{eq:PLV}), which also gives the posterior samples of PLV averaged over $n$ subjects. 
Based on the posterior samples of PLV, one can obtain posterior mean of PLV as a point estimate and $95\%$ credible intervals for measure of uncertainty quantification. 
Moreover, one can also compute the posterior probability that PLV is greater than a specific value (e.g. 0.7).
This enables uncertainty-aware analysis regarding whether PLV exceeds a given threshold. 
For example, one may consider an edge to be connected if the posterior probability of PLV exceeding the threshold is greater than 0.5, thereby incorporating uncertainty into network construction and avoiding overconfident decisions based on noisy estimates.

\section{Numerical Experiments with EEG Data}\label{sec:num}

Here we demonstrate the proposed functional circular models  through real EEG dataset from \cite{chennu2016brain}. 
In particular, by using a real dataset and adding artificial noise to the original data, we evaluate how the proposed method-based PLV is noise-robust and the standard PLV is not.  
We used resting-state human EEG dataset (250 Hz sampling, 2500 time points) collected on 91 electrodes from 20 participants \citep{chennu2016brain}. 
Following \cite{chennu2016brain}, we obtained alpha (8Hz--15Hz) and beta (12Hz--25Hz) components by bandpass filters and applied the Hilbert transform (MATLAB function \textit{hilbert}) to compute their phases. 
Then, we extracted the first $T=100$ samples.
Thus, we obtained $p=91$ dimensional functional data (taking values on unit circle) for each participant.

\begin{figure}[htbp!]
\centering
\includegraphics[width=\linewidth]{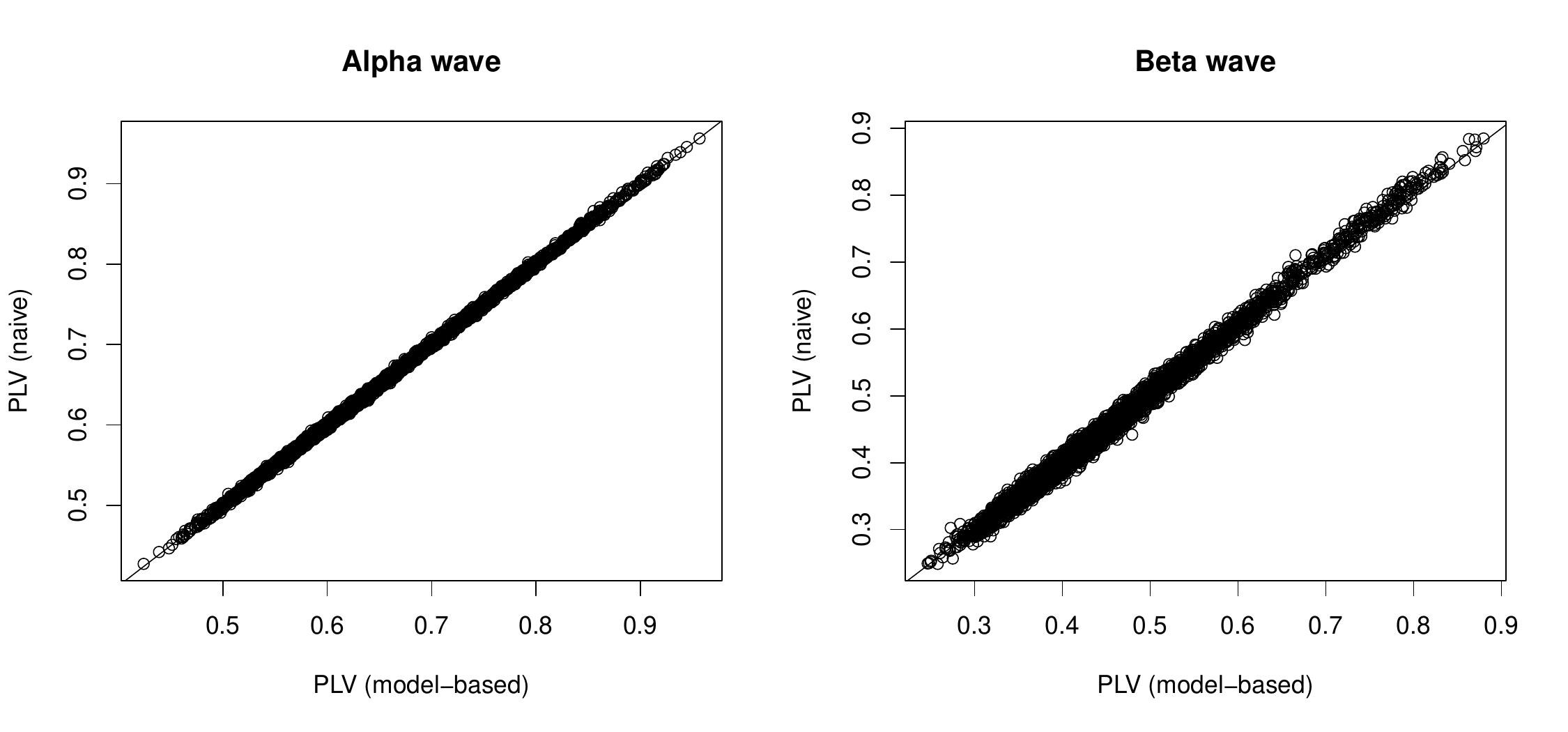}\\
\includegraphics[width=0.45\linewidth]{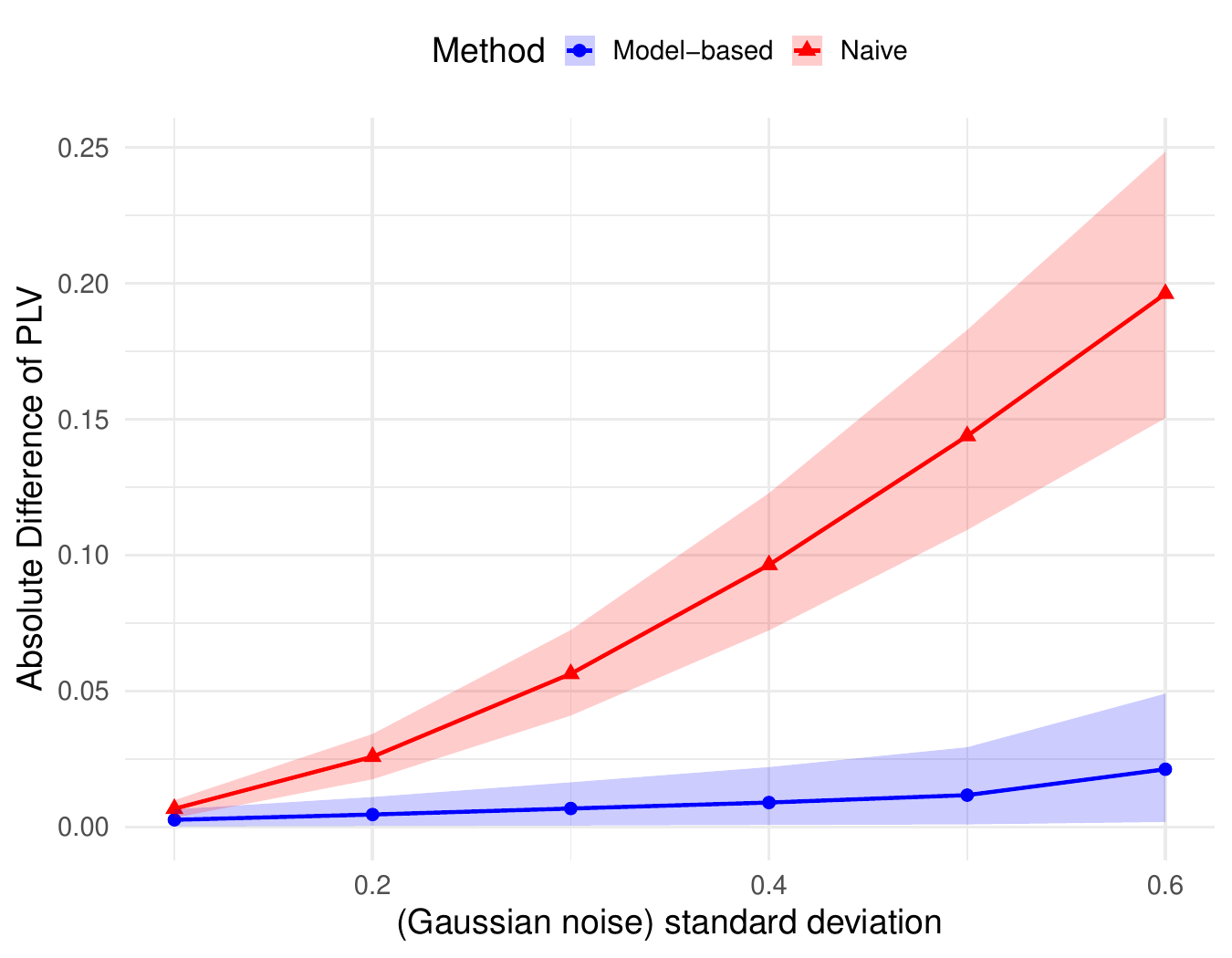}
\includegraphics[width=0.45\linewidth]{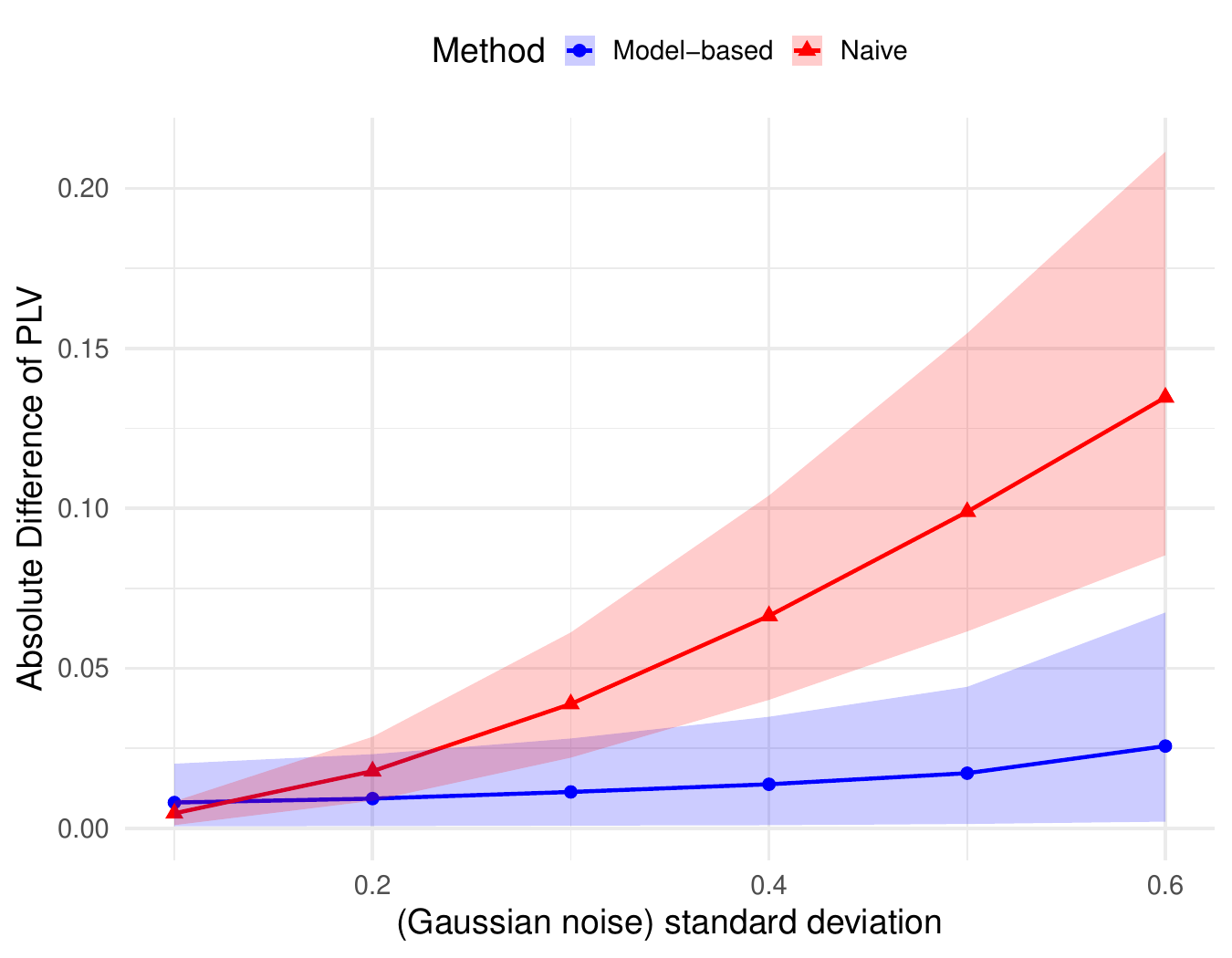}\\
\includegraphics[width=0.45\linewidth]{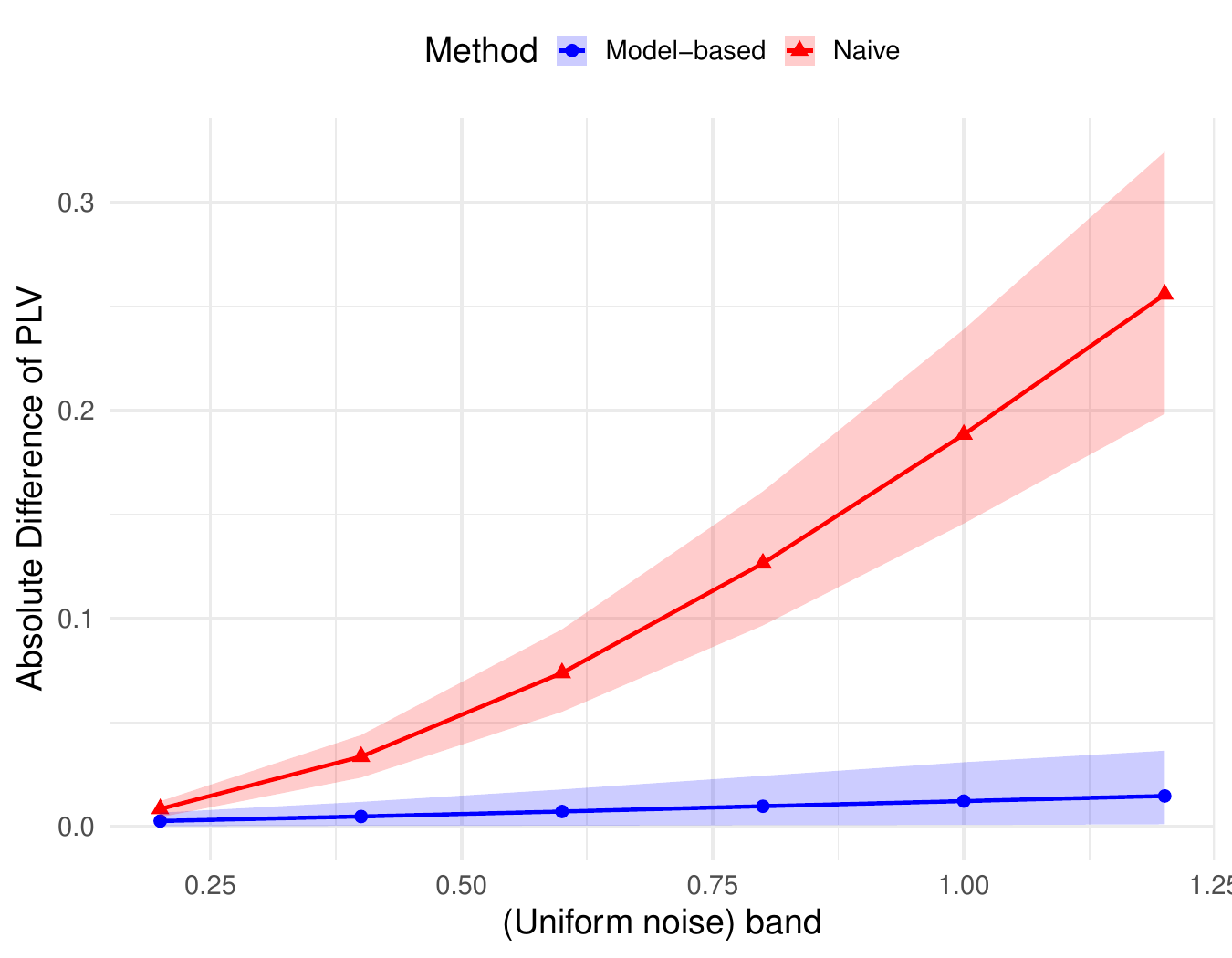}
\includegraphics[width=0.45\linewidth]{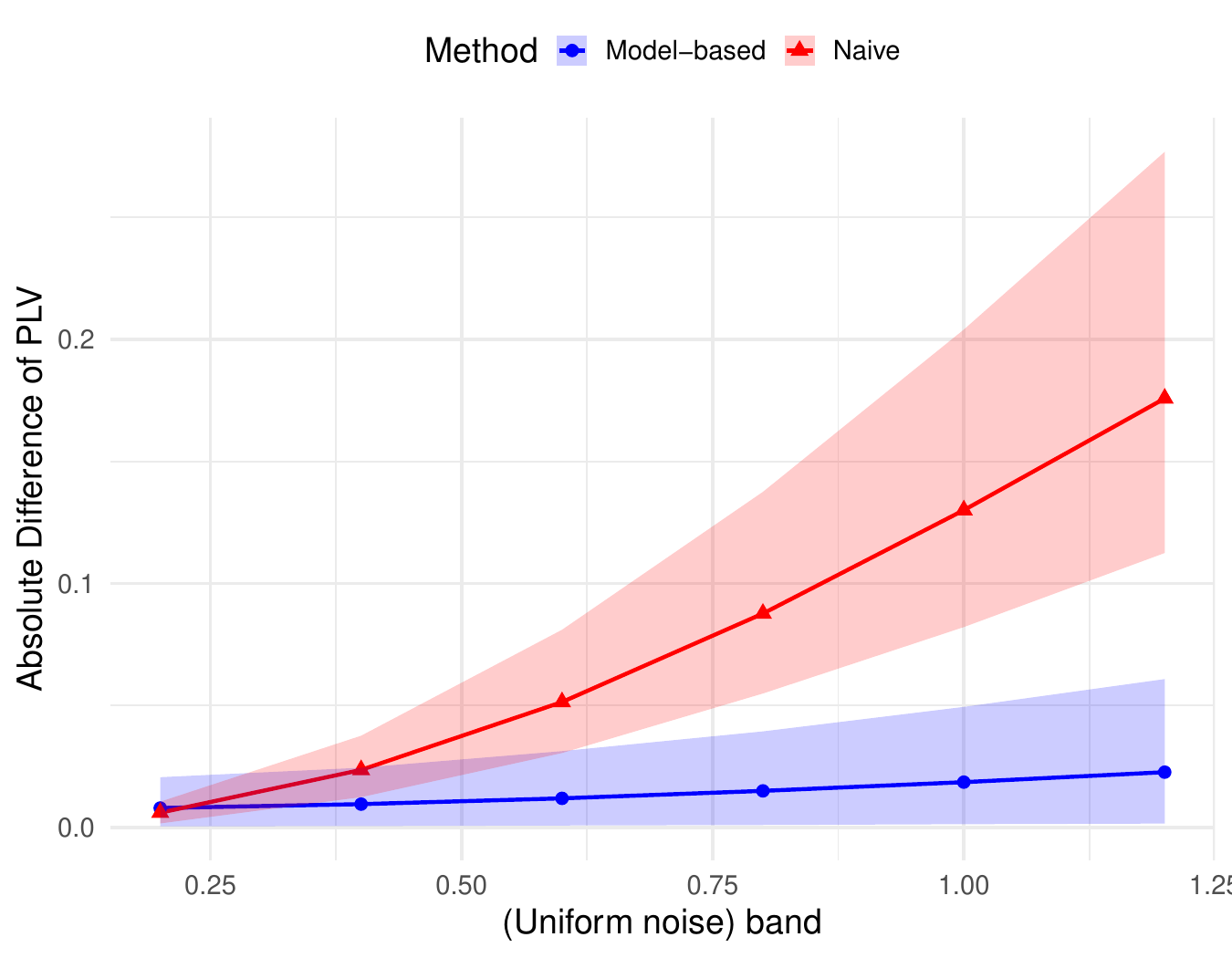}
\caption{ Top: Scatter plots of model-based and naive PLV values obtained from the original data. 
Bottom: Average values of absolute difference between estimates (under noisy data) and grand truth (under noise-less data) among $p(p-1)/2=4005$ relationships, based on EEG alpha (left) and beta (right) wave data. 
The colored region corresponds to point-wise upper and lower $5\%$ quantiles of the absolute differences.    } 
\label{fig:app}
\end{figure}

For the dataset, we first calculated PLV values based on the observed data and applied the proposed wrapped functional model (WFM) to compute the model-based PLV.  
For the proposed method, we generated 1000 posterior samples after discarding the first 1000 samples as burn-in.
In the top panel of Figure~\ref{fig:app}, we show the scatter plots of the naive PLV (based on observation) and model-based PLV, showing that the two methods provide similar estimates in both alpha and beta waves. 
We set the estimates as the grand truth and evaluate how the noise affects the estimation performance by adding artificial noise of varying types and magnitudes to the original dataset.
Specifically, we considered Gaussian noise, $N(0, b^2)$ and uniform noise, $U(-b, b)$ where $b$ controls the strength of noise. 
For six noise levels $b\in\{0.1, 0.2,\ldots,0.6\}$ for Gaussian noise and $b\in \{0.2, 0.4,\ldots,1.2\}$ for uniform noise, we generated noised data as $\widetilde{Y}_{kj}^{(s)}=Y_{kj}^{(s)}+\ep_{kj}^{(s)}\  ({\rm mod}\  2\pi)$, where $Y_{kj}^{(s)}$ is the original observation and $\ep_{kj}^{(s)}$ is the generated noise. 
Based on the noisy observation $\widetilde{Y}_{kj}^{(s)}$, we calculated the naive PLV and model-based PLV via CFM.
For CFM, we generated 1000 posterior samples after discarding the first 1000 samples as burn-in, and computed posterior means of PLV as point estimates. 
The point estimates of PLV are compared with the grand truth via absolute difference.
We then summarized the results by reporting the 
average absolute difference across all pairs, together with the point-wise 5\% upper and 
lower quantiles at different noise levels.
The results are displayed in Figure~\ref{fig:app}. 
It is clearly seen that the naive PLV is highly sensitive to noise contamination and the naive estimates quickly deviate from the true PLVs even with relatively small noise levels.
On the other hand, the model-based PLV remains stable and robust and the accuracy of the proposed method is well preserved even as the noise intensity is 
gradually increased, highlighting its noise-robust properties and advantage of using CFM for de-noising.

We next assessed the utility of PLV for network construction by 
evaluating whether each PLV exceeds a pre-specified threshold. We adopted 0.7 as a cutoff value, and for each method we conducted a binary classification analysis to distinguish whether PLV is larger than 0.7. 
For the CFM-based PLV, we calculated the posterior probability that PLV is greater than 0.7, namely, $p_{k,k'}\equiv {\rm P}(\overline{{\rm PLV}}_{k,k'}\geq 0.7 \mid {\rm Data})$, based on the posterior samples of PLV, and regarded as significantly large than 0.7 when $p_{k,k'}\geq 0.5$.
To evaluate the performance of binary classification, we calculated the true positive rate (TPR) and F1 score under various noise levels. 
The results are summarized in Table~\ref{tab:sim}. 
The results show that the proposed CFM-based method achieves consistently high TPRs and F1 scores across different conditions, while the naive PLV exhibits severe deterioration. In particular, when the noise reaches a moderate level, the naive PLV completely loses the ability to detect edges above the 0.7 threshold, resulting in TPR values close to zero. 
By contrast, the proposed CFM method maintains accurate classification performance, confirming its robustness.

\begin{figure}[htbp!]
\centering
\includegraphics[width=0.45\linewidth]{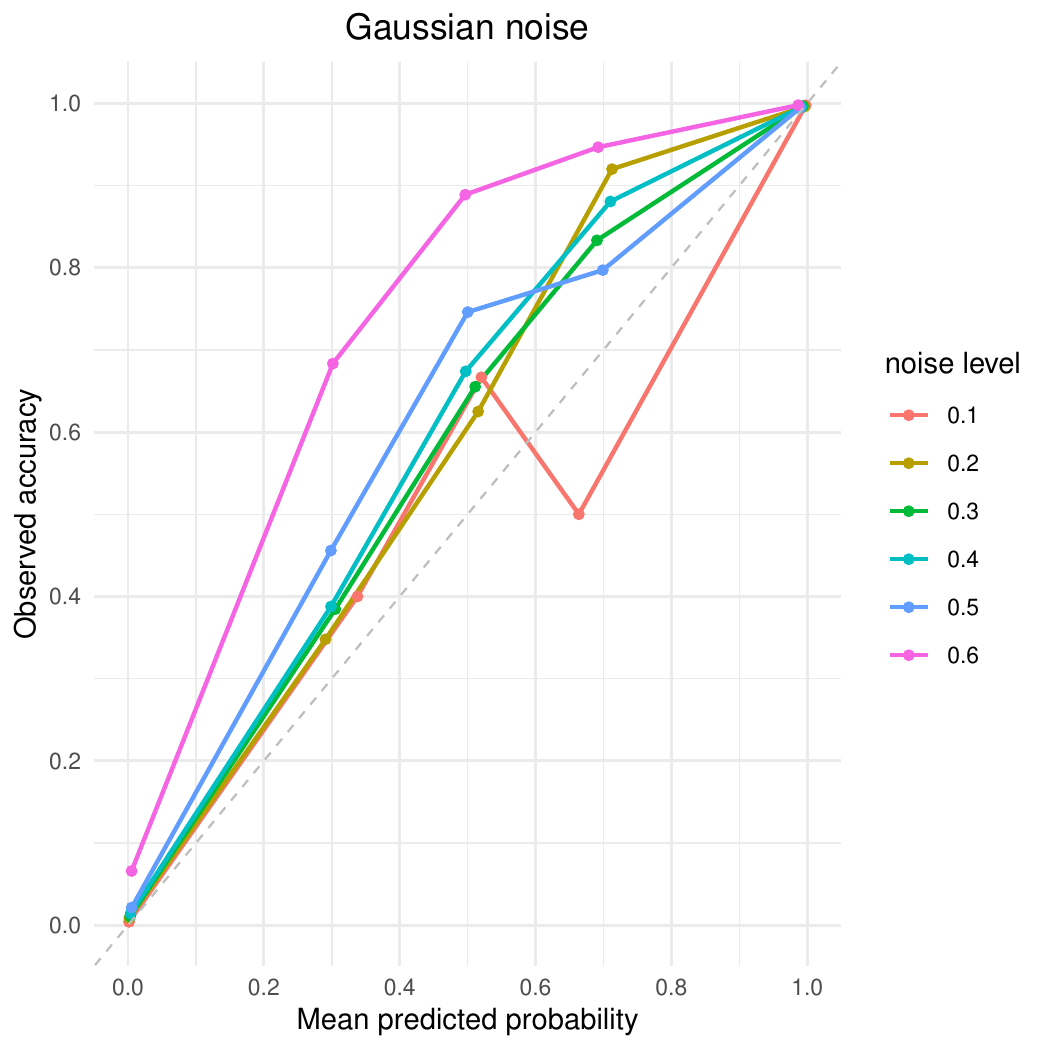}
\includegraphics[width=0.45\linewidth]{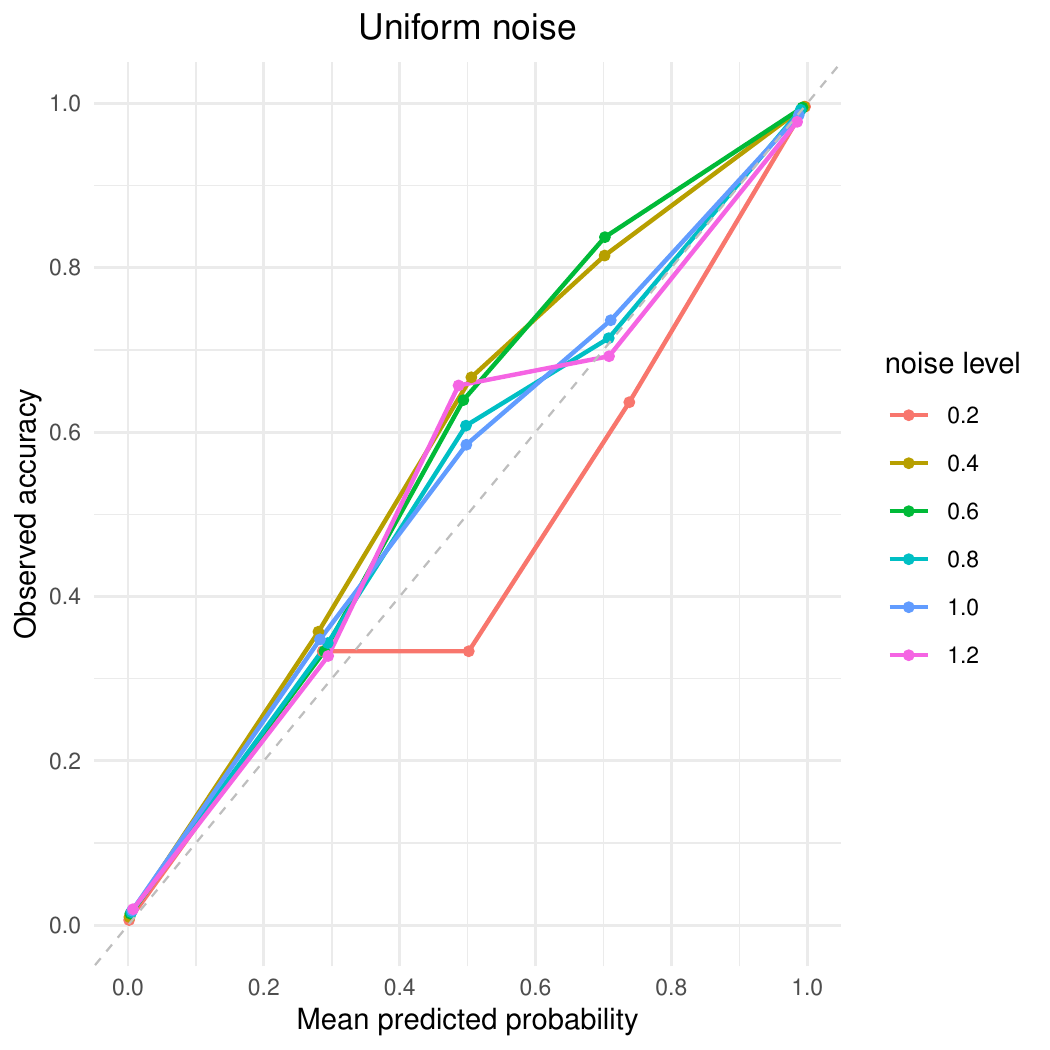}
\caption{ Scatter plots of posterior probability and prediction accuracy under Gaussian (left) and uniform (right) noise. } 
\label{fig:app-reliability}
\end{figure}

Finally, we examined whether the Bayesian posterior distributions produced by the proposed method provide reliable uncertainty quantification. 
Using the posterior probability $p_{k,k'}$ for each pair $(k, k')$, we first grouped these posterior probabilities into five categories with intervals, $[0, 0.2]$, $[0.2, 0.4],\ldots, [0.8, 1]$.
For each group, we calculated the empirical frequency that the true PLV actually exceeds 0.7 and the average values of the posterior probability $p_{k,k'}$ within the same group. 
In Figure~\ref{fig:app-reliability}, the scatter plots of the empirical frequency and average posterior probability are shown under alpha wave with different noise levels. 
It is observed that the scatter plots align closely with the 45-degree line, indicating that the posterior probabilities from CFM are well calibrated, that is, if the posterior probability is, for example, 0.8, then indeed about 80\% of those edges exceed the true threshold. 
This demonstrates that the proposed CFM method not only produces accurate point estimates of PLV but also delivers meaningful and trustworthy uncertainty assessments even under noisy data.

\vspace{1cm}

\begin{table}[htb!]
\caption{
True positive rates (TPR) and F1 values for significant edges with PLV larger than 0.7, obtained by circular functional model (CFM) and direct calculation of PLV from the observed data.  
\label{tab:sim}}
\centering

\medskip
\begin{tabular}{cccccccccccc}
\hline
 &&& \multicolumn{4}{c}{Alpha wave} && \multicolumn{4}{c}{Beta wave} \\
 &&& \multicolumn{2}{c}{TPR (\%)} & \multicolumn{2}{c}{F1 ($\times 100$)} && \multicolumn{2}{c}{TPR (\%)} & \multicolumn{2}{c}{F1 ($\times 100$)}\\
 & Level &  & CFM & Direct & CFM & Direct &  & CFM & Direct & CFM & Direct \\
\hline
 & 0.1 &  & 98.6 & 93.8 & 98.9 & 96.8 &  & 90.7 & 94.0 & 94.9 & 96.9 \\
 & 0.2 &  & 97.4 & 73.3 & 98.3 & 84.6 &  & 89.4 & 78.2 & 94.4 & 87.8 \\
{\footnotesize Gaussian} & 0.3 &  & 96.2 & 44.7 & 97.5 & 61.8 &  & 83.3 & 52.3 & 90.9 & 68.7 \\
{\footnotesize noise} & 0.4 &  & 94.5 & 17.8 & 96.5 & 30.2 &  & 79.6 & 13.0 & 88.7 & 23.0 \\
 & 0.5 &  & 91.0 & 3.1 & 94.2 & 6.1 &  & 71.8 & 0.0 & 83.6 & 0.0 \\
 & 0.6 &  & 80.0 & 0.0 & 88.5 & 0.0 &  & 49.5 & 0.0 & 66.3 & 0.0 \\
\hline
 & 0.2 &  & 98.1 & 91.2 & 98.7 & 95.4 &  & 89.4 & 92.6 & 94.1 & 96.2 \\
 & 0.4 &  & 96.5 & 65.1 & 97.7 & 78.9 &  & 86.6 & 71.8 & 92.8 & 83.6 \\
{\footnotesize Uniform} & 0.6 &  & 95.0 & 30.7 & 96.8 & 46.9 &  & 81.5 & 31.9 & 89.8 & 48.4 \\
{\footnotesize noise} & 0.8 &  & 94.1 & 6.0 & 95.8 & 11.3 &  & 74.1 & 1.9 & 85.1 & 3.6 \\
 & 1 &  & 92.7 & 0.0 & 94.5 & 0.0 &  & 63.0 & 0.0 & 77.3 & 0.0 \\
 & 1.2 &  & 91.1 & 0.0 & 93.1 & 0.0 &  & 54.2 & 0.0 & 70.3 & 0.0 \\
\hline
\end{tabular}
\end{table}

\vspace{1cm}

\section{Concluding remarks}\label{sec:conc}
This paper introduced a Bayesian framework for noise-robust inference on PLV using circular functional modeling (CFM).
Unlike conventional PLV, which is sensitive to noise and lacks uncertainty quantification, the proposed CFM model enables denoising and provides reliable posterior inference. 
Through numerical experiments with real EEG data, the method demonstrated stable performance under noise and well-calibrated uncertainty assessment based on posterior probabilities of exceeding a threshold.

Since the EEG data typically entails information of observed locations, such information could be incorporated into the proposed model through, for example, spatially correlated random effects through, for example, Gaussian process in the CFM method.  
However, such approach may require substantial computational costs especially when the number of locations is large. 
While the proposed method does not explicitly take account of such information, it would be a reasonable compromised approach in terms of computational cost and estimation accuracy.

\section*{Acknowledgement}
This work is partially supported by JSPS KAKENHI Grant Numbers 24K21420 and 25H00546.
Takeru Matsuda was supported by JSPS KAKENHI Grant Numbers 19K20220, 21H05205, 22K17865 and JST Moonshot Grant Number JPMJMS2024.

\appendix

\section{Markov Chain Monte Carlo algorithm via Gibbs sampler}
We provide details of posterior computation algorithm to fit the proposed multi-level circular functional model. 
The joint posterior of $\sigma^2, \Theta=\{ (\tau_l, \beta_l, \gamma_l), \ l=1,\ldots,L\}$, the random coefficient $M=\{a_{kl}^{(s)}, \mu_{kl}, \ \ s=1,\ldots,n,\ k=1,\ldots,p,\ l=1,\ldots,L\}$ and the latent variable $Z=\{Z_{kj}^{(s)},\ s=1,\ldots,n, \ k=1,\ldots,p,\ j=1,\ldots,T\}$ given the observed data $\mathcal{D}=\{Y_{kj}^{(s)},\ s=1,\ldots,n, \ k=1,\ldots,p,\ j=1,\ldots,T\}$ can be expressed as 
\begin{equation*}
\begin{split}
&p(\sigma^2, \Theta, M, Z \mid \mathcal{D}) \\
&\propto 
\exp\left\{-\frac1{2\sigma^2}\sum_{s=1}^n \sum_{k=1}^p \sum_{j=1}^T \Big(Y_{kj}^{(s)} - \sum_{l=1}^L a_{kl}^{(s)}B_{lj}+ 2\pi Z_{kj}^{(s)}\Big)^2\right\}p(\sigma^2) (\sigma^2)^{-npT/2}\\
&  \times p(\Theta) \prod_{l=1}^L  (\tau_l^2)^{-np/2} (\gamma_l^2)^{-p/2}\exp\left\{-\frac{1}{2\tau_l^2}\sum_{s=1}^n \sum_{k=1}^p (a_{kl}^{(s)}-\mu_{kl})^2 - \frac{1}{2\gamma_l^2}\sum_{k=1}^p (\mu_{kl}-\beta_l)^2\right\},
\end{split}
\end{equation*}
where $p(\Theta)$ and $p(\sigma^2)$ are prior distributions for $\Theta$ and $\sigma^2$, respectively.
Sampling from the joint posterior can be easily done by a simple Gibbs sampling, where the details of each sampling step are provided as follows: 

\begin{itemize}
\item[-] 
(Sampling from $a_k^{(s)}\equiv (a_{k1}^{(s)},\ldots,a_{kL}^{(s)})$) \ 
For $s=1,\ldots,n$ and $k=1,\ldots,p$, generate $a_k^{(s)}$ from its full conditional distribution, $N(\widetilde{A}^{-1}\widetilde{B}_k^{(s)}, \widetilde{A}^{-1})$, where 
$$
\widetilde{A}= \frac{1}{\sigma^2} \sum_{j=1}^T B_jB_j^\top + D_\tau , \ \ \ \ \ 
\widetilde{B}_k^{(s)}= \frac{1}{\sigma^2} \sum_{j=1}^T B_j (Y_{kj}^{(s)}+2\pi Z_{kj}^{(s)}) + D_\tau \widetilde{M}_k,
$$
where $B_j=(B_{1j},\ldots, B_{Lj})^\top$, $D_\tau={\rm diag}(1/\tau_1^2,\ldots, 1/\tau_L^2)$, and $\widetilde{M}_k=(\mu_{1k},\ldots,\mu_{Lk})^\top$.

\item[-] 
(Sampling from $\mu_{kl}$) \ For $k=1,\ldots,p$ and $l=1,\ldots,L$, generate $\mu_{kl}$ from its full conditional distribution as follows:  
$$
N\left(
\left(\frac{n}{\tau_l^2} + \frac{1}{\gamma_l^2}\right)^{-1}\left(\frac{1}{\tau_l^2}\sum_{s=1}^n a_{kl}^{(s)} + \frac{\beta_l}{\gamma_l^2}\right),
\left(\frac{n}{\tau_l^2} + \frac{1}{\gamma_l^2}\right)^{-1}
\right).
$$

\item[-] 
(Sampling from $\beta_l$) \ For $l=1,\ldots,L$, generate $\beta_l$ from its full conditional distribution as follows: 
$$
N\left( \left( \frac{p}{\gamma_l^2} + \frac{1}{B_0}\right)^{-1} \left(\frac{1}{\gamma_l^2}\sum_{k=1}^p \mu_{kl} + \frac{A_0}{B_0}\right), 
\left( \frac{p}{\gamma_l^2} + \frac{1}{B_0}\right)^{-1}
\right)
$$

\item[-] 
(Sampling from $\tau_l^2$) \ For $l=1,\ldots,L$, generate $\tau_l^2$ from its full conditional distribution as follows: 
$$
{\rm IG}\left( 
\nu_\tau + \frac{np}{2},\  
\eta_\tau + \frac12\sum_{i=1}^n \sum_{k=1}^p (a_{kl}^{(s)}-\mu_{kl})^2\right).
$$

\item[-] 
(Sampling from $\gamma_l^2$) \ For $l=1,\ldots,L$, generate $\gamma_l^2$ from its full conditional distribution as follows: 
$$
{\rm IG}\left( 
\nu_\gamma + \frac{p}{2},\  
\eta_\gamma + \frac12\sum_{i=1}^n \sum_{k=1}^p (\mu_{kl}-\beta_l)^2\right).
$$

\item[-] 
(Sampling from $\sigma^2$) \ Generate $\sigma^2$ from its full conditional distribution ${\rm IG}(\widetilde{\nu}_\sigma,\widetilde{\nu}_\sigma)$, where $\widetilde{\nu}_\sigma=\nu_\sigma+npT/2$ and 
$$
\widetilde{\eta}_\sigma=\eta_\sigma + \frac12\sum_{s=1}^n \sum_{k=1}^p \sum_{j=1}^T \Big(Y_{kl}^{(s)} - \sum_{l=1}^L a_{kl}^{(s)}B_{lj}+ 2\pi Z_{kj}^{(s)}\Big)^2.
$$

\item[-] 
(Sampling from $Z_{kj}^{(s)}$) \ For $i=1,\ldots,n$, $k=1,\ldots,p$ and $j=1,\ldots,T$, the full conditional distribution of $Z_{kj}^{(s)}$ is proportional to $\phi(Y_{kj}^{(s)}-\sum_{l=1}^L a_{kl}^{(s)}B_{lj}; -2\pi Z_{kj}, \sigma^2)$, where $\phi(x; a, b)$ denotes the normal density with mean $a$ and variance $b$.  
Then, we first set a natural number $\tilde{M}$ such that the density $\phi(Y_{kj}^{(s)}-\sum_{l=1}^L a_{kl}^{(s)}B_{lj}; -2\pi Z_{kj}, \sigma^2)$ is sufficiently small for all $|Z_{kj}^{(s)}|>\tilde{M}$. 
Then, generate $Z_{kj}^{(s)}$ from the multinomial distribution on $\{0, \pm 1,\ldots, \pm \tilde{M}\}$ with probability of $Z_{kj}^{(s)}=m$ given by 
$$
P(Z_{kj}^{(s)}=m)=\frac
{\phi(Y_{kj}^{(s)}-\sum_{l=1}^L a_{kl}^{(s)}B_{lj}; -2\pi m, \sigma^2)}
{\sum_{m'=-\tilde{M}}^{\tilde{M}} \phi(Y_{kj}^{(s)}-\sum_{l=1}^L a_{kl}^{(s)}B_{lj}; -2\pi m', \sigma^2)},
$$
for $m=0, \pm 1, \ldots, \pm \tilde{M}$.
\end{itemize}

\vspace{1cm}
\bibliographystyle{chicago}
\bibliography{ref}

\end{document}